\documentclass[authoryear]{article}
\usepackage[]{graphicx}\usepackage[]{color}
\usepackage{alltt} 

\usepackage[left=3cm,right=3cm, top = 2.5cm, bottom = 2.5cm]{geometry}

\usepackage{natbib}
\usepackage[utf8]{inputenc}
\usepackage[UKenglish]{babel}
\usepackage[T1]{fontenc}
\usepackage{amsmath}
\usepackage{amssymb}
\usepackage{xcolor}

\usepackage{url}

\newcommand{\DAIC}{\Delta\mbox{AIC}}
\newcommand{\eg}{\textit{e.g.}\ }
\newcommand{\ie}{\textit{i.e.}\ }
\DeclareMathOperator{\NegBin}{NBin}
\newcommand{\given}{\,\vert\,}

\usepackage[normalem]{ulem}
\newcommand{\old}[1]{}

\IfFileExists{upquote.sty}{\usepackage{upquote}}{}
\begin{document}

\title{Endemic-epidemic models with
discrete-time serial interval distributions for infectious disease prediction}

\author{Johannes Bracher and Leonhard Held}

\maketitle

\begin{abstract}
Multivariate count time series models are an important tool for the
analysis and prediction of infectious disease spread. We
consider the endemic-epidemic framework, an autoregressive model
class for infectious disease surveillance counts, and replace the default
autoregression on counts from the previous time period with more
flexible weighting schemes inspired by discrete-time serial interval
distributions. We employ three different parametric formulations,
each with an additional unknown weighting parameter estimated via a profile
likelihood approach, and compare them to an unrestricted
nonparametric approach.  The new methods are illustrated in a
univariate analysis of dengue fever incidence in San Juan, Puerto
Rico, and a spatio-temporal study of viral gastroenteritis in the
twelve districts of Berlin. We assess the predictive performance of
the suggested models and several reference models at various forecast horizons.
In both applications, the performance of the endemic-epidemic models is considerably
improved by the proposed weighting schemes.
\end{abstract}


\section{Introduction}

Infectious disease surveillance produces multivariate time series of
counts, usually available on a weekly basis. Models for such data can
help to understand mechanisms of spread or to predict future incidence.
As \citet{Keeling2008} point out, the different goals are often in
conflict. While prediction requires high accuracy, transparency of
models is important to better understand disease spread. This
trade-off has given rise to a wide spectrum of methods
\citep{Siettos2013}, from individual-level to compartmental
to statistical approaches.

In this
article we consider the endemic-epidemic (in the following: EE)
framework, a class of statistical time series models for multivariate surveillance
counts introduced by \citet{Held2005} and extended in a series of
articles \citep{Paul2008, Held2012, Meyer2014, Meyer2017a}.
In its current formulation and
implementation in the \texttt{R} package
\texttt{surveillance} \citep{Meyer2017} the EE framework uses only incidence from the directly preceding week $t - 1$ to explain incidence in week $t$. In a mechanistic interpretation of such a \textit{first-order} model, the time between the appearance of symptoms in successive generations
is thus assumed to be fixed to the observation interval at which the data are collected, here one week. However, in reality this time span, called the {\em serial interval} \citep[p.156]{Becker2015}, varies randomly across infection events and may exceed one observation interval. Other factors like incomplete reporting or external drivers may likewise introduce dependencies at larger time horizons, which the EE model cannot currently capture. In this paper we address this limitation by introducing more flexible weighting schemes for past incidence inspired by different shapes of the serial interval distribution. Specifically we
consider shifted Poisson, triangular and geometric weights and compare them to an unrestricted nonparametric approach.

Univariate EE models with geometric lag weights have
close links to integer-valued GARCH (INGARCH) time series models
\citep{Fokianos2009, Zhu2011}. Multivariate INGARCH
models have also been considered in the time series literature in recent
years \citep{Heinen2007, Cui2018, Fokianos2020}. However, they do not address the particular challenges encountered in
infectious disease epidemiology, as they are not able to describe the
spatio-temporal spread of infectious diseases \citep{Xia2004} or to
accommodate social contact data \citep{Mossong2008}. On a
practical note, no general implementation of multivariate INGARCH models
seems to be available, whereas the methods presented in this paper can be applied
by practitioners using the \texttt{R} packages \texttt{surveillance} and
\texttt{hhh4addon} (see \ref{app:software}). Recent applications to malaria and
cutaneous leishmaniasis \citep{Adegboye2017}, dengue \citep{Cheng2016, Zhu2019} and
invasive pneumococcal disease \citep{Chiavenna2019} illustrate the practical usefulness
of the EE framework.

Forecasting is a key task in infectious disease epidemiology and may
help to guide interventions, but has proven to be
challenging \citep{Moran2016}. In recent years the topic has received
increasing attention due to large-scale forecasting competitions
such as the CDC \textit{FluSight} Challenge \citep{McGowan2019}, the
DAPRA Chikungunya Challenge \citep{DelValle2018}, the
NOAA Dengue Forecasting Project \citep{PPFSTIWG2015},
and the RAPIDD Ebola Forecasting Challenge \citep{Viboud2018}.
While \eg the \textit{FluSight} Challenges also feature forecasts for multiple geographical sub-units (the 10 Health and Human Services Regions of the US), the different forecasts are evaluated and often generated separately in a univariate fashion. The EE framework, in contrast, has been
repeatedly used for joint spatio-temporal count prediction \citep{Paul2011,
Meyer2014,Held2017}, and general methodology to evaluate such forecasts has
been provided in \cite{Held2017} and \cite{Held2019}. Other approaches
for spatio-temporal epidemic forecasting include deterministic compartmental
models \citep{Yang2016, Pei2018}, statistical matching \citep{Viboud2003}
and spline methods \citep{Bauer2016}. An overview on spatio-temporal
epidemic modelling, also covering
time-series SIR models \citep{Xia2004}, can be found in \cite{Wakefield2019}.

In two case studies we examine how the suggested weighting schemes
can improve the predictive performance of EE models. The
first considers weekly dengue counts in San Juan, Puerto
Rico. \citet{Ray2017} recently used these data to compare the
forecasting performance of their kernel conditional density estimation
(KCDE) method to a first-order EE model and found
their approach to give better results. A possible reason is that the
KCDE method conditions forecasts on several preceding observations
rather than just the most recent. In the second case study we assess the benefits
of the suggested model extension in multivariate EE models
for viral gastroenteritis in the twelve districts of Berlin.
Here we propose to evaluate multivariate forecasts at various horizons via
multivariate logarithmic scores \citep{Gneiting2008} and provide an
efficient Monte Carlo algorithm for computation. For comparison we consider forecasts from na\"ive seasonal models and negative binomial GLARMA (generalized linear autoregressive moving average) models \citep{Dunsmuir2015}.

The remainder of the article is structured as follows. Section \ref{sec:case_studies} introduces the
two case studies. In Section \ref{sec:methodology} we extend the EE
model class by including weighting schemes for past incidence and give details on inference
and prediction. Section \ref{sec:applications} presents the results from our case studies,
before we conclude with a discussion in Section \ref{sec:discussion}.

\section{Data}
\label{sec:case_studies}

\subsection{Incidence of dengue fever in San Juan, Puerto Rico}
\label{subsec:dengue_data}

\begin{figure}[h!]
\includegraphics[width=\textwidth]{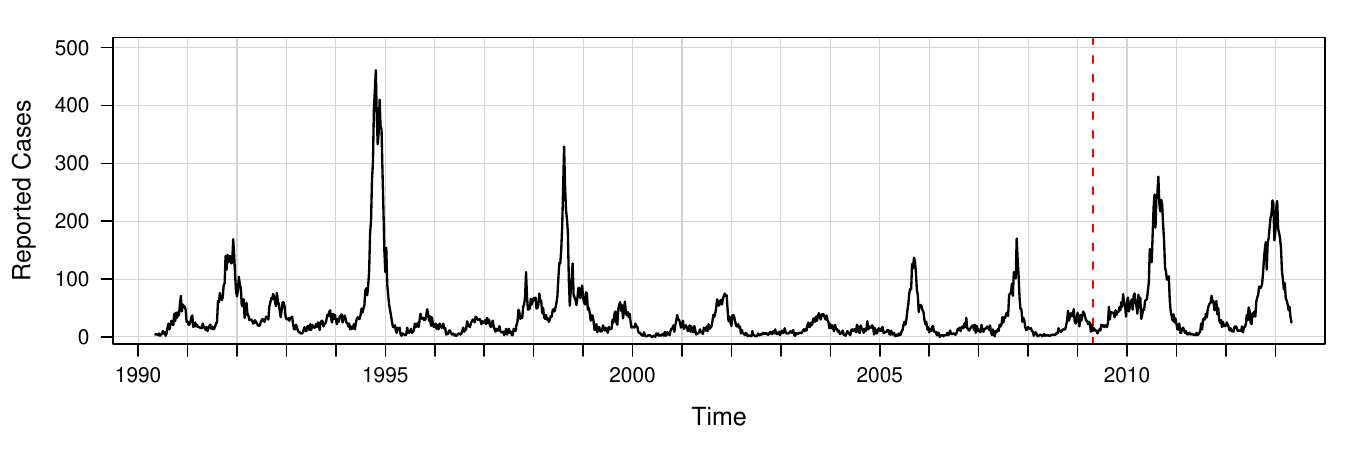} 
\caption{Weekly numbers of reported dengue cases in San Juan, Puerto Rico, 1990--2013. The dashed line marks the boundary between training and test data.}
\label{fig:dengue}
\end{figure}

Weekly counts of reported dengue cases in San Juan, Puerto Rico from
week 1990-W18 through 2013-W17 are shown in Figure
\ref{fig:dengue}. These have previously been analysed by
\citet{Ray2017} following the NOAA Dengue Forecasting Project
\citep{PPFSTIWG2015}. Dengue, a viral febrile illness transmitted by
mosquitoes, is endemic in most tropical and subtropical regions
\citep{Heymann2015}. Incidence is highest during summer and early autumn.
The incubation period is 4--7 days while the mean
serial interval is estimated at 15--17 days
\citep{Aldstadt2012}. To put this into perspective, we note that this mean serial interval is considerably longer than for seasonal influenza (2--3 days), similar to varicella (14 days, \citealt{Vink2014}), and much shorter than for \eg tuberculosis (estimates between 6 and 18 months, \citealt{Ma2018}). As in
\citet{Ray2017} the data are split into a training
(1990-W18--2009-W17, 988 weeks) and a test period (2009-W18--2013-W17,
208 weeks) where the forecast performance is assessed.

\subsection{Viral gastroenteritis incidence in Berlin, Germany}
\label{subsec:case_study_berlin}

\begin{figure}[h!]
\center
\includegraphics[width=0.92\textwidth]{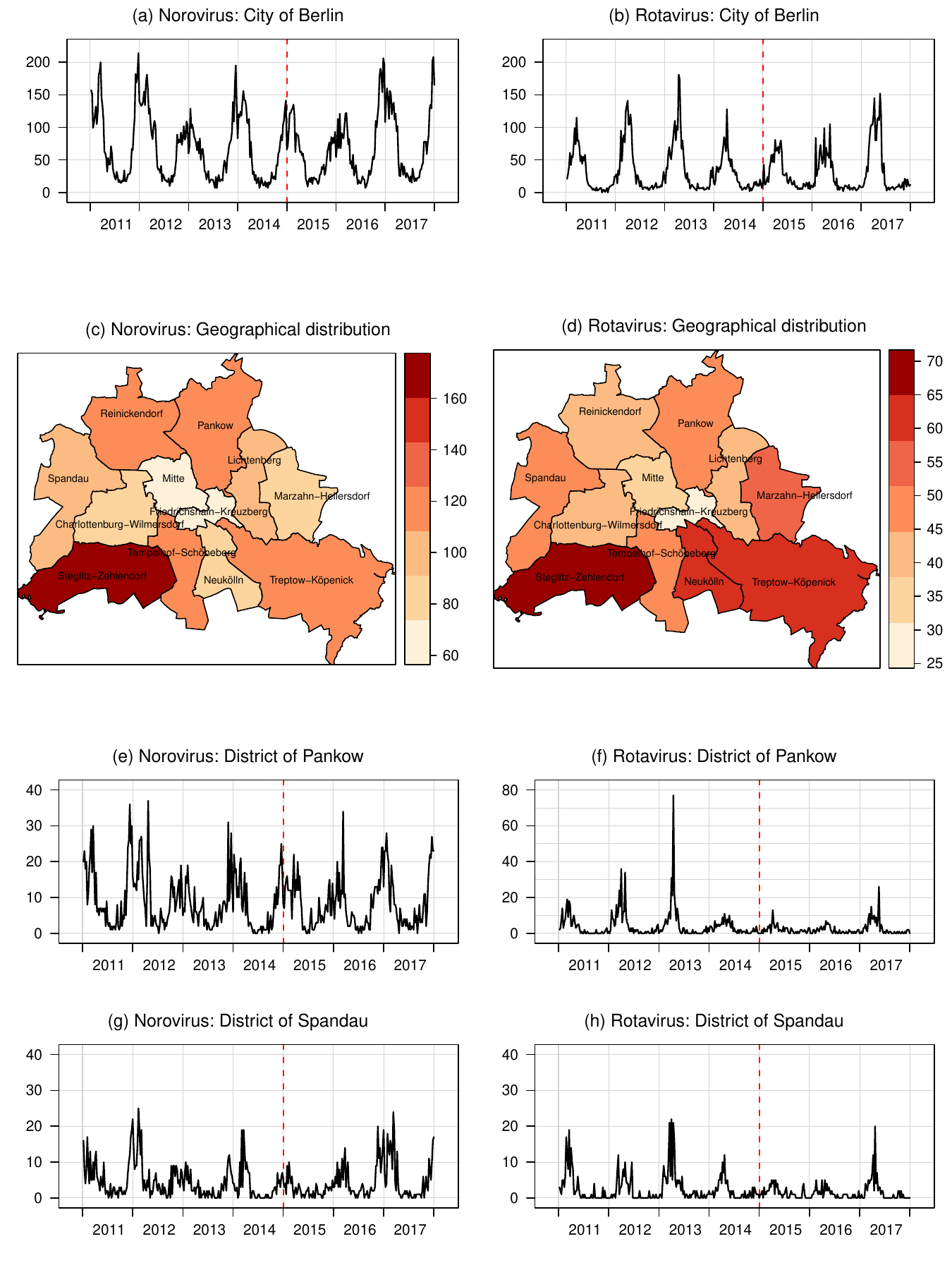} 
\caption{Weekly numbers of reported norovirus (left) and rotavirus (right) cases in Berlin, 2011--2017. Top row: Pooled over the twelve city districts. Second row: Average number of cases per year and 100,000 inhabitants in the twelve districts. Third and fourth row: Districts of Pankow and Spandau. Dotted lines show the beginning of the test period. Plots for the remaining districts can be found in the Supplementary Material.}
\label{fig:incidence}
\end{figure}

Norovirus and rotavirus are common causes of viral gastroenteritis and
can be transmitted from person to person or via
contaminated food and items. The average serial interval
of norovirus is 3--4 days \citep{Richardson2001}, but the virus may be
shed for up to three weeks after symptom resolution \citep{Heymann2015}.
For rotavirus the average serial interval is 5--6 days \citep{Richardson2001},
with infectiousness lasting for up to 12 days \citep{CDC2015}. For rotavirus a vaccine has been available since 2006, while this is not the case for norovirus. In
Germany both diseases are
notifiable 
and the Robert Koch Institute makes weekly numbers of reported cases available
({\url{https://survstat.rki.de}}). We consider counts
from the twelve districts of Berlin from week 2011-W01
through 2017-W52 (downloaded on 30 Aug 2018). For both diseases we will assess the predictive performance over the weeks 2015-W01 through 2017-W52 (156 weeks) based on at least four years (208 weeks) of training data. Figure
\ref{fig:incidence} shows the observed incidence aggregated to city-wide weekly counts
(panels a and b) and the spatial distribution across the twelve
districts (panels c and d). For illustration we also show counts from Pankow and Spandau, Berlin's largest and smallest districts, respectively (panels e--h; see the Supplementary Material for the other districts).
Note that parts of the norovirus data have been previously analysed with first-order EE models \citep{Meyer2017a, Held2017}.

\section{Methodology}
\label{sec:methodology}

\subsection{The endemic-epidemic model}
\label{subsec:endemic_epidemic}

We provide a brief overview on the current state of the endemic-epidemic (EE) framework. Given the past, the number of cases $Y_{it}$ in unit $i = 1, \dots, m$ and week $t$ is assumed to follow a negative binomial distribution
\begin{align}
Y_{it} \given \mathbf{Y}_{t - 1}, \mathbf{Y}_{t - 2}, \dots & \sim \NegBin(\lambda_{it}, \psi_i)\label{eq:neg_bin}
\end{align}
with conditional mean $\lambda_{it}$ and overdispersion parameter
$\psi_i$. The conditional variance is thus
$\lambda_{it} + \psi_i\lambda_{it}^2$ and reduces to $\lambda_{it}$ for $\psi_i=0$.
A common simplification is to assume one global overdispersion
parameter, \ie $\psi_i = \psi$ for $i = 1, \dots, m$. Counts from
different units $i$ and $j$ at time $t$ are assumed to be conditionally independent given the past.

The conditional mean $\lambda_{it}$ in \eqref{eq:neg_bin} is further decomposed into
\begin{align}
\lambda_{it} & = \nu_{it} + \phi_{it} \sum_{j = 1}^m \lfloor w_{ji}\rfloor Y_{j, t - 1},\label{eq:hhh4_mean_structure}
\end{align}
where $\nu_{it}$ is referred to as the \textit{endemic component} and captures infections not directly linked to observed cases from the previous week. The remaining autoregressive term in \eqref{eq:hhh4_mean_structure} is the \textit{epidemic component} and describes how the incidence in region $i$ is linked to previous cases $Y_{j, t - 1}$ in units $j = 1, \dots, m$. The parameters $\nu_{it}$ and $\phi_{it}$ are constrained to be non-negative and modelled in a log-linear fashion, for instance with sine-cosine terms to account for seasonality \citep{Held2012}. Long-term temporal trends and covariates such as meteorological conditions \citep{Cheng2016, Bauer2018} or vaccination coverage  \citep{Herzog2011} could also be included.

The coupling between units is described by weights $w_{ji}$, which enter
in \eqref{eq:hhh4_mean_structure} after normalisation:
$\lfloor w_{ji}\rfloor = w_{ji}/\sum_{h = 1}^m
w_{jh}$. Unrestricted estimation of all $m^2$ weights is usually
unstable in practice, so more parsimonious and epidemiologically
meaningful parameterizations have been introduced. Specifically, the weights can be based on
social contact data for spread across age
groups \citep{Meyer2017a} or on the geographical distance between cases to describe spatio-temporal spread \citep{Meyer2014}. In the latter case the weights can be specified through a power law
\begin{align}
w_{ji} & = (o_{ji} + 1)^{-\rho}\label{eq:power_law}
\end{align}
where $o_{ji}$ is the path distance between the regions $j$ and $i$ (with $o_{ii} = 0$, $o_{ji} = 1$ for direct neighbours $i$ and $j$ and so on) and $\rho$ is a decay parameter to be estimated from the data. The power law formulation is motivated from human movement behaviour \citep{Brockmann2006} and has been found to be an efficient way of capturing spatial dependence.

\subsection{Weighting schemes for past incidence}
\label{subsec:higher_order_lags}

The EE framework was originally formulated as an observation-driven statistical time series model \citep{cox:1981}, but a direct derivation from a discrete-time susceptible-infectious-removed (SIR) model has recently been provided by \citet{Bauer2018} and \citet{Wakefield2019}. They derive formulation \eqref{eq:hhh4_mean_structure} from a competing risks framework where the forces of infection from the infected individuals in units $j = 1, \dots, m$ on a susceptible in unit $i$ add up to an overall risk of infection. An important assumption leading to equation \eqref{eq:hhh4_mean_structure} is that infectiousness lasts for exactly one time period, and infection at time $t$ is only possible by cases from $t - 1$. In other words, the serial interval is fixed to one observation interval. If this assumption is violated, a simple remedy is to aggregate the data to a time scale corresponding to the average serial interval, as done by \citet{Herzog2011} and \citet{Wakefield2019} for measles. This, however, still does not account for the fact that serial intervals are random in reality and vary across infection events. This can be included in the model via an autoregression on several past observations, where the (discretized) distribution of the serial intervals is reflected in the weights assigned to the different lags (\citealt{ForsbergWhite2008}, \citealt{Becker2015}, p.156).

Moreover, it has been shown that underreporting can introduce dependencies at higher lags, even if the underlying process is a first-order autoregression \citep{Bracher2019}. External drivers such as meteorological conditions may have similar effects. This means that even for diseases with typically short serial intervals, including higher-order lags may yield improved predictive performance. At the same time, confounding by such factors implies that there may not always be direct correspondence between the estimated lag weights and the serial interval distribution.

We start by considering univariate EE model with time-constant endemic and epidemic parameters $\nu$ and $\phi$, respectively, which we extend to
\begin{equation}\label{eq:simpleModel.u}
\lambda_t = \nu + \phi \sum_{d=1}^p \lfloor u_d \rfloor Y_{t - d}.
\end{equation}
Here, the weights $\lfloor u_d \rfloor = u_d/\sum_{g = 1}^p u_g$ are normalized and restricted to be positive. This is in line with the motivation from a discrete-time serial interval distribution (where $\lfloor u_d \rfloor$ is the probability for a serial interval of $d$ weeks), but also has technical reasons. The sum-to-one constraint is required for identifiability while positivity of the weights ensures well-definedness of the model (allowing $u_d < 0$ can result in negative $\lambda_t$, in which case the negative binomial distribution is not defined).

Estimating the normalized weights $\lfloor u_d\rfloor$ without further parametric assumptions is the most flexible approach, but the large number of additional parameters may lead to numerical instabilities and suboptimal predictive performance. The other
extreme is to work with a pre-specified weighting scheme based on epidemiological knowledge. For instance \cite{Wang2011} fix the weights at $u_1 = 5/6, u_2 = 1/6$ in their study on hand, mouth and foot disease, which is considered a plausible description of the infectious period. As a flexible compromise between these two approaches we will moreover consider parametric weighting schemes corresponding to different shapes of the (discrete-time) serial interval distribution \citep[p.156]{Becker2015}. This is combined with data-driven estimation of an underlying scalar parameter.

Specifically, we employ the following three parameterizations, always
truncated to lags $d = 1, \dots, p$ and depending on an unknown weighting
parameter $\kappa$. \textit{Shifted Poisson} weights
correspond to
\begin{equation}
u_d = \frac{\kappa^{d - 1}}{(d - 1)!}\exp(-\kappa), \ \ \kappa > 0, \label{eq:pois_weights}
\end{equation}
as used by \citet{Kucharski2014} to analyse daily avian influenza data. A \textit{linear decay} of the  weights (subject to a non-negativity constraint),
\begin{equation}
u_d = \max(1 - \kappa d, 0)\label{eq:lin_weights}, \ \ 0 <\kappa < 1,
\end{equation}
corresponds to a \textit{triangular} serial interval distribution. Finally, a \textit{geometric} weighting scheme
\begin{equation}
u_d = (1 - \kappa)^{d - 1}\kappa, \ \ 0 < \kappa < 1, \label{eq:geom_weights}
\end{equation}
is considered. The latter can also be motivated by the impact underreporting has on the autocorrelation of a first-order autoregressive process \citep{Bracher2019}. Moreover note that the formulation \eqref{eq:simpleModel.u} with geometric weights \eqref{eq:geom_weights} and $p \rightarrow \infty$
is equivalent to the negative binomial INGARCH(1, 1) model
\begin{equation}
\lambda_t = \alpha + \beta Y_{t - 1} + \gamma \lambda_{t - 1},\label{eq:link_ingarch}
\end{equation}
where $\alpha = \nu\kappa, \beta = \phi\kappa$ and $\gamma = 1 - \kappa$ \citep{Fokianos2009,Zhu2011}.

Parameterizations \eqref{eq:lin_weights} and \eqref{eq:geom_weights} force the weight $u_1$ to be the largest and imply a decay in infectiousness. This is reasonable for weekly data and diseases with short serial intervals. The Poisson version \eqref{eq:pois_weights} can also capture an initial increase in infectiousness and may thus be more appropriate for longer serial intervals or daily data.
For comparison we will also consider the current \textit{first-order} EE formulation where all weight is concentrated on the first lag,
\begin{equation}
u_1 = 1, u_2 = \ldots = u_p = 0.\label{eq:fixed}
\end{equation}

We may also combine the multivariate formulation
\eqref{eq:hhh4_mean_structure} with the suggested weighting schemes,
\begin{align}
\lambda_{it} & = \nu_{it} + \phi_{it} \sum_{d = 1}^p \sum_{j = 1}^m \lfloor u_d\rfloor \lfloor w_{ji} \rfloor Y_{j, t - d}\label{eq:multiv_distr_lag2},
\end{align}
where $\nu_{it}$ and $\phi_{it}$ now depend on unit $i$ and time $t$.
The weights $\lfloor u_d\rfloor$, however, are assumed to be constant across units and time.

\subsection{Inference}
\label{subsec:inference}

We have extended the likelihood-based inference procedure from the \texttt{R} package \texttt{surveillance} \citep{Meyer2017} to fit higher-order lag models \eqref{eq:pois_weights}--\eqref{eq:geom_weights} in the package \texttt{hhh4addon} (see \ref{app:software}). To optimize the likelihood given the weights $\lfloor u_d \rfloor$, the efficient and robust routine provided in \texttt{surveillance} for first-order models \citep{Paul2008,Paul2011,Meyer2014} has been adapted. To facilitate integration with the existing implementation, the weighting parameter $\kappa$ (suitably transformed to the real line) or the unrestricted weights $\lfloor u_d \rfloor$ (on a multinomial logit scale) are estimated via a profile likelihood approach \citep{Held2014}. This way, there is no need to implement cumbersome derivatives of the log likelihood function with respect to the additional parameters. Note that a similar approach has previously been applied to fit EE models with an additional age stratification \citep{Meyer2017}.

The maximum of the profile likelihood function can be found using standard numerical optimization. If the weights are estimated in an unrestricted nonparametric manner it may be necessary to try several starting values to ensure convergence. This is also the case for the triangular parameterization \eqref{eq:lin_weights}, which sometimes leads to several local optima (see remarks in Section \ref{subsec:model_fits}). 

Under the parametric formulations \eqref{eq:pois_weights}--\eqref{eq:geom_weights},
the weights $\lfloor u_d\rfloor$ usually become negligible after a certain order. To find a sufficiently large $p$ we suggest to  inspect the weights visually and to plot the order $p$ against the AIC (or log-likelihood,
as the complexity of the model does not depend on $p$) to check when changes become negligible.
For the unrestricted estimation of the weights $\lfloor u_1\rfloor,
\dots, \lfloor u_p \rfloor$, where the number of model parameters depends on $p$, we choose the order based on the
AIC, as is common for prediction purposes.

\subsection{Prediction and predictive model assessment}

In order to take into account the uncertainty surrounding epidemiological forecasts, these should take the form of predictive distributions rather than deterministic point forecasts \citep{Held2017}. We consider weekly updated forecasts for different horizons $h = 1, \dots, H$, as in the recent CDC \textit{FluSight} challenge \citep{McGowan2019}. This means that for each week $t$, the model is re-fitted using all data already available in order to predict the incidence in weeks $t + 1, \dots, t + H$. As in previous work \citep{Held2019} we use plug-in forecasts, \ie forecast from the fitted model without carrying forward the uncertainty in the parameter estimates. Taking this into account would require fitting EE models using MCMC methods in a Bayesian framework \citep{Bauer2018, Wakefield2019} with carefully chosen prior distributions.

Plug-in one-week-ahead forecast distributions from EE models are given by independent negative binomial distributions for the $m$ units, so the predictive densities can be evaluated analytically. For forecast horizons of two or more weeks ahead, predictive means, variances and autocovariances can be still obtained analytically in the EE class, see the Supplementary Material for details. The full predictive distributions, however, are no longer given by any standard distribution and are dependent across different units, as they reflect \eg spatio-temporal dependencies arising over a longer forecast horizon. To evaluate the predictive probability masses $f(\mathbf{Y}_{t + h} \given M)$ we apply the following procedure, which uses Rao-Blackwellization \citep{GelfandSmith1990,Ray2017} for improved numerical stability:

\begin{enumerate}
\item Generate $k = 1, \ldots, K$ samples from the predictive distribution of \\ $(\mathbf{Y}_{t + 1}, \dots, \mathbf{Y}_{t + h - 1} \given M)$.
\item For each sample $(\mathbf{y}_{t + 1}^{(k)}, \dots \mathbf{y}_{t + h - 1}^{(k)})$ and each unit $i = 1, \dots, m$ evaluate
$$
f(Y_{i, t + h} \given  M, \mathbf{y}_{t + 1}^{(k)}, \dots, \mathbf{y}_{t + h - 1}^{(k)})
$$
on a suitably chosen support $y_{i, t + h} \in \{0, \dots, Y_{i, t + h}^{(\text{max})}\}$. These one-week-ahead distributions are conditionally independent negative binomial distributions. The joint probability mass function $f(\mathbf{Y}_{t + h} \given  M, \mathbf{y}_{t + 1}^{(k)}, \dots, \mathbf{y}_{t + h - 1}^{(k)})$ is thus given by the product of the unit-wise probability mass functions.
\item Compute $f(\mathbf{Y}_{t + h} \given M) = 1/K \cdot \sum_{k = 1}^K f(\mathbf{Y}_{t + h} \given  M, \mathbf{y}_{t + 1}^{(k)}, \dots, \mathbf{y}_{t + h - 1}^{(k)})$.
\end{enumerate}
As the relevant support of $\mathbf{Y}_{t + h}$ obviously gets very large even for moderate $m$, we usually only store the unit-wise marginal predictive densities or summaries thereof (central 50\% and 95\% predictive intervals), supplemented with analytically obtained covariances. In all applications we set $K = 1000$, which led to very stable results.

To assess the performance of probabilistic forecasts, proper scoring rules \citep{Gneiting2007} have become the standard in infectious disease epidemiology \citep{Held2017}. A score is called proper if there is no incentive for a forecaster to digress from her true belief, and strictly proper if any such digress leads to a penalty. A commonly used strictly proper score is the logarithmic score
$$
\text{logS}(y_{\text{obs}} \given M) = -\log[f(y_{\text{obs}} \given M)],
$$
\ie the negative log predictive probability mass the model $M$ assigned to the observed value $y_{\text{obs}}$. The score is negatively oriented, \ie lower values are better. To aggregate over several forecasts one can use average log scores, which are still strictly proper. For multivariate $h$-week ahead forecasts of $\mathbf{Y}_{t + h}$ we apply the multivariate logarithmic score \citep{Gneiting2008}
\begin{equation}\label{eq:MultLS}
\text{logS}(\mathbf{y}_{t + h} \given M) = -\log[f(\mathbf{y}_{t + h} \given M)]
\end{equation}
where $M$ is now a model fitted based on the information available
at time $t$. In practice the score \eqref{eq:MultLS} is often standardized and
divided by $m = \text{dim}(\mathbf{y}_{t})$.
For one-week-ahead predictions from EE models, the standardized multivariate log score can be evaluated analytically as
it is just the average of $m$ negative binomial log-densities. For $h \geq 2$ we apply the simulation-based approach described above.

\subsection{Reference forecasts}
\label{subsec:reference_models}

Throughout this manuscript we will compare our methods to two other forecasting approaches. The first is a na\"ive seasonal model which does not take into account any serial dependence. It is given by a negative binomial generalized linear model (GLM) including the calendar week and, in the multivariate case, district as categorical covariates (fitting separate models per district was not possible as the frequent zero counts in our data led to degenerate forecast distributions). The second reference model is the negative binomial GLARMA (generalized linear autoregressive moving average) model implemented in the \texttt{R} package \texttt{glarma} \citep{Dunsmuir2015}. In contrast to the EE model, the GLARMA model also allows for negative dependencies between observations. As in \cite{Held2019} we fitted this model with the same sine-cosine terms for seasonality as the respective EE models. Details on the GLARMA model are provided in the Supplementary Material.


\section{Applications}
\label{sec:applications}

\subsection{Incidence of dengue fever in San Juan, Puerto Rico}
\label{sec:application_dengue}

\citet{Ray2017} compare the predictive performance of the EE model framework with their kernel conditional density estimation (KCDE) method. They apply the following first-order EE model to the weekly dengue counts described in Section \ref{subsec:dengue_data}:
\begin{align}
\lambda_t & = \nu_t + \phi_t Y_{t - 1} \label{eq:lambda_dengue}\\
\log(\nu_{t}) & = \alpha^{(\nu)} + \gamma^{(\nu)}\sin(2\pi t/\omega) + \delta^{(\nu)}\cos(2\pi t/\omega)\label{eq:nu_dengue}\\
\log(\phi_{t}) & = \alpha^{(\phi)} + \sum_{k = 1}^2 \gamma_k^{(\phi)}\sin(2\pi k t/\omega) + \delta_k^{(\phi)}\cos(2\pi k t/\omega), \label{eq:phi_dengue}
\end{align}
where $\omega=52$ to model yearly seasonality in weekly data.
This formulation has been chosen by \cite{Ray2017} from a number of
candidate models based on the AIC on the training data (1990-W18
through 2009-W17, see Figure \ref{fig:dengue}). Note that unlike
in other works on dengue \citep{Cheng2016, Chen2019},
\citet{Ray2017} do not incorporate meteorological data. To ensure
comparability with \citet{Ray2017} we only replace
\eqref{eq:lambda_dengue} with \eqref{eq:simpleModel.u} but stick to
the decomposition of $\log(\nu_t)$ and $\log(\phi_t)$ as in
\eqref{eq:nu_dengue} and \eqref{eq:phi_dengue}. For the weights
$\lfloor u_d \rfloor$ we apply unrestricted nonparametric estimation
and the parametric formulations
\eqref{eq:pois_weights}--\eqref{eq:geom_weights}. In addition, we use a weighting scheme based on an estimate of the serial interval distribution from \citeauthor{Siraj2017} (\citeyear{Siraj2017}, Fig. 1). Discretized to a weekly scale, it is given by $u_1 = 0, u_2 = 0.2, u_3 = 0.425, u_4 = 0.25, u_5 = 0.125$. On a standard laptop the model fitting takes
around one second per model for the parametric versions and four seconds for the unrestricted one. We did
not re-run the KCDE forecasts from \cite{Ray2017}, since their extensive Supplementary
Material (\url{https://github.com/reichlab/article-disease-pred-with-kcde})
allowed us to compute all quantities required in our comparison.
However, previous
studies \citep{Held2019} indicate that KCDE requires at least an order
of magnitude more computation time than the EE approach.

\subsubsection{Model fit and residual correlation}

\begin{figure}[h!]
\includegraphics[width=\textwidth]{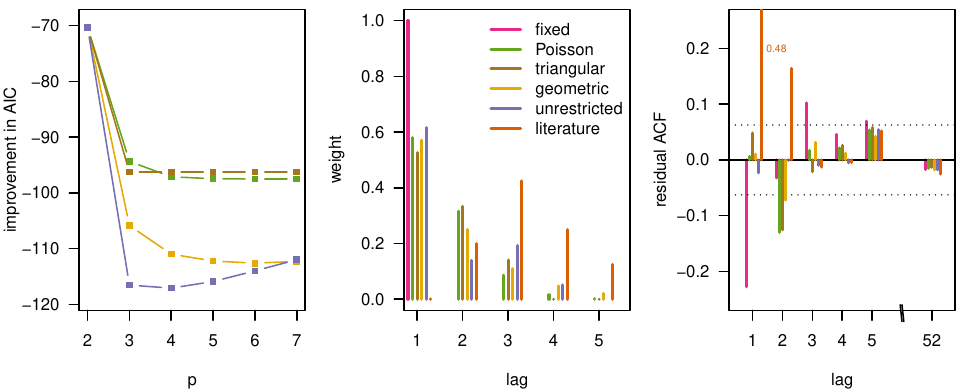} 


\caption{Left: Improvement in AIC (relative to the first-order EE model) for the different lag weight parameterizations and values of $p$, applied to dengue counts 1990--2009. Middle: Weights $\lfloor u_d \rfloor$ (with $p = 4$ in the unrestricted case, $p = 5$ otherwise). Right: Autocorrelations of Pearson residuals. The model denoted by \textit{literature} is based on the serial interval distribution from \cite{Siraj2017}.}
\label{fig:lag_weights_dengue}
\end{figure}

Figure \ref{fig:lag_weights_dengue}, left panel, shows the AIC values achieved for the training data under the different parameterizations of the lag weights and orders $p$ 
relative to the first-order EE model (AIC = 6671.1) used in \citeauthor{Ray2017} (\citeyear{Ray2017}; we omit comparison to the KCDE method here as no AIC values are available). The largest improvement ($\DAIC=-117.1$) is achieved by a model with $p = 4$ and unrestricted weights $\lfloor u_d\rfloor$. For all parametric versions \eqref{eq:pois_weights}--\eqref{eq:geom_weights} the AIC stays practically constant from $p = 5$ on, with the geometric version performing better ($\DAIC=-112.2$) than the Poisson and triangular versions ($\DAIC=-97.5$ and $-96.3$, respectively). The model using the serial interval distribution from \citeauthor{Siraj2017} (\citeyear{Siraj2017}, referred to as \textit{literature} in the legend) is omitted here as it has a considerably higher AIC ($\Delta \text{AIC} = 186$ relative to the first-order model).

The weights $\lfloor u_d \rfloor$ resulting from the different parameterizations are shown in the middle panel of Figure \ref{fig:lag_weights_dengue} (with $p = 4$ for the nonparametric version and $p = 5$ otherwise). The unrestricted weights show a non-monotonic pattern where the third lag is more important than the second. The parametric weights all imply decaying weights, with the geometric version assigning more weight to lags 4 and 5 than two than the Poisson and triangular ones. These results are in contrast to the weights based on the serial interval distribution from \cite{Siraj2017}, indicating that other factors like meteorological conditions also impact the autocorrelation structure and thus the estimated lag weights. Plots of the profile likelihood for each parameterization can be found in the Supplementary Material.

The autocorrelation functions of the Pearson residuals $(y_{t} - \lambda_{t})/\sqrt{\lambda_t + \psi\lambda_t^2}$ shown in the right panel of Figure \ref{fig:lag_weights_dengue} look unsuspicious for the nonparametric and the geometric versions. The simpler first-order model suffers from strong negative autocorrelation at lag 1, while for the Poisson and triangular serial interval distributions these occur at lag 2. These patterns are in line with the fact that relative to the nonparametric version these parameterizations assign too much weight to lags one and two, respectively. The model with pre-specified weights based on the serial interval shows excessive residual autocorrelations at lag one, likely because it does not use the first lag and fails to capture short-time dependencies induced by external factors.

\subsubsection{Predictive assessment}

We now assess the predictive performance of the different EE models and compare them to the KCDE approach
\citep{Ray2017} and the reference models over the course of the seasons 2009/10 through
2012/13. Unlike \citet{Ray2017}, who averaged scores over
prediction horizons $1, \dots, 52$, we focus on short-term forecasts and assess performance separately for the horizons $h = 1, \dots, 8$ weeks. For the KCDE method the respective average scores (over the 208 weeks of test data) could be computed from material provided in the Supplement of \cite{Ray2017}.

\begin{figure}
\includegraphics[width=\textwidth]{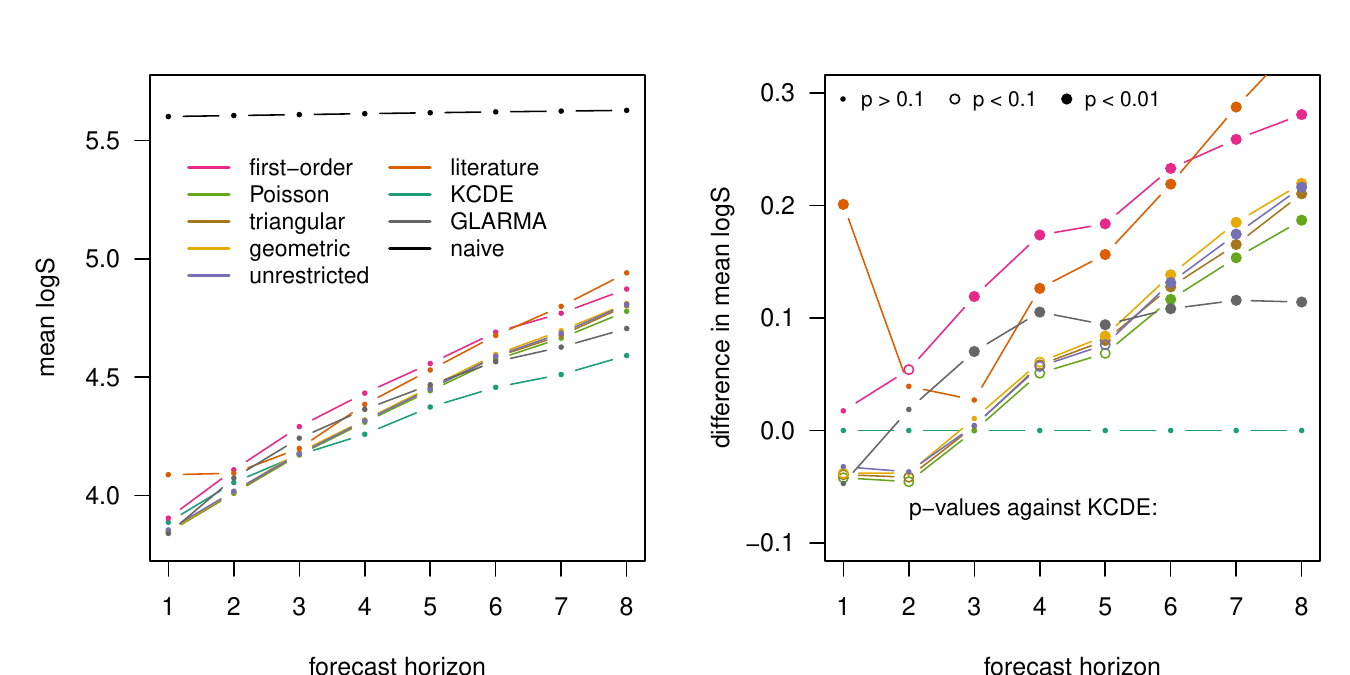} 
\caption{Left: average log scores of dengue forecasts at horizons $h = 1, \dots, 8$. Right: differences in average log scores compared to KCDE (negative values indicate better performance of EE methods). The point shape in the right plot indicates whether there is evidence for different forecasting performance of KCDE and the respective EE method (based on two-sided $p$-values from permutation tests; \citealt{Paul2011}). All numbers underlying this figure can be found in the Supplementary Material.}
\label{fig:log_scores_dengue}
\end{figure}

The left panel of Figure \ref{fig:log_scores_dengue} shows the
average scores obtained by the periodic, full bandwidth KCDE method,
the six variants of the EE model, and the two reference models from Section \ref{subsec:reference_models}. The weighting based on the serial interval distribution from \cite{Siraj2017} leads to slight improvements relative to the first-order model for horizons 2 through 6, but considerably worse performance of one-week-ahead forecasts. The EE models with data-driven weighting schemes yield better performance than the first-order version at all horizons, while differences between the four parameterizations are minor.
Notably, the unrestricted nonparametric approach
does not lead to a performance gain compared to the parametric versions.
For one- and two-week ahead predictions the extended EE models
also outperform the KCDE method. The differences,
shown in detail in the right panel of Figure \ref{fig:log_scores_dengue},
may be spurious, however. Permutation tests \citep{Paul2011}
indicate only weak evidence that the forecasts from the extended EE
models actually perform better (two-sided $p$-values between
0.05 and
0.18 for $h=1$ and
0.06 and 0.13 for $h=2$). While for $h = 3$ performance is almost identical, from $h = 4$
onwards there is moderate to strong evidence that KCDE performs better
($p$-values of 0.04
and smaller; see Supplementary Material). A possible reason is that
KCDE uses separate models optimized for prediction at different horizons,
while the EE method generates all forecasts iteratively from the same model.
Concerning the reference models, the na\"ive seasonal model is clearly outperformed by all other approaches. The GLARMA model performs very similarly to the higher-order EE models for one-week-ahead forecasts, somewhat worse for horizons 2 through 4 and somewhat better for horizons 6 through 8 (but worse than KCDE).

\begin{figure}
\includegraphics[width=\textwidth]{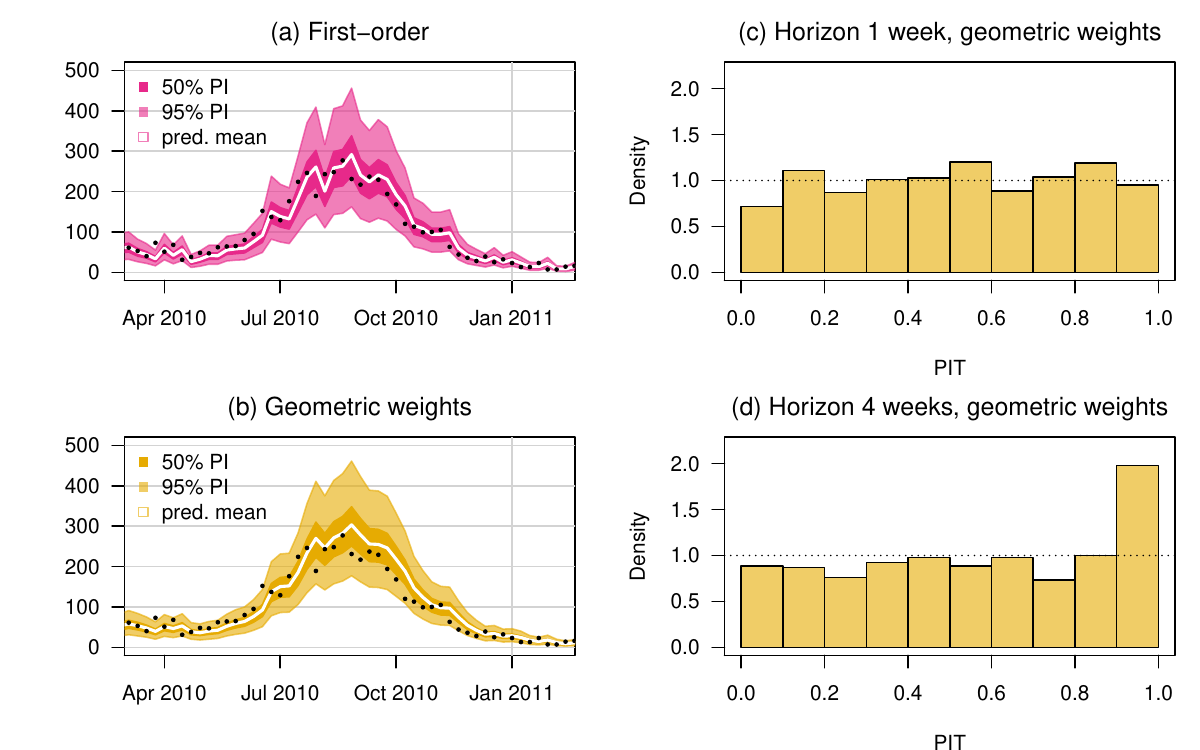} 
\caption{Left: One-week ahead 50\% and 95\% prediction intervals for dengue, season 2010/2011, obtained from EE models with (a) only first lag and (b) geometric weights. Right: PIT histograms of (c) one-week-ahead and (d) four-week-ahead forecasts from the model with geometric weights.}
\label{fig:forecasts_dengue}
\end{figure}

Visualizations of forecasts from different models and at different horizons can be found in the Supplementary Material. We here restrict ourselves to a comparison of the central 50\% and 95\% prediction intervals from the first-order and geometric lag EE models, shown in the left panel of Figure \ref{fig:forecasts_dengue}. As they are based on an average of past values rather than just the last, the predictions from the model with geometric weights form a smoother curve. The right panel of the figure moreover shows probability integral transform (PIT) histograms of one-week-ahead and four-week ahead forecasts from the model with geometric weights, \ie the distribution of $\text{PIT}_{t + h} = \text{Pr}(Y_{t + h} \leq y_{t + h} \given M)$ (with a correction for the discreteness of the data, \citealt{Czado2009}). If forecasts are well calibrated, the PIT histogram should be approximately uniform, as is the case for a forecast horizon of one week. At a horizon of four weeks, there are too many large PIT values, \ie observations towards the upper tail of the respective predictive distribution. Note, however, that strictly speaking PIT histograms should be applied to a set of mutually independent forecasts, which is not the case for horizons $h > 1$.

\subsection{Viral gastroenteritis incidence in Berlin, Germany}
\label{sec:application_berlin}

We now apply multivariate EE models to the data on norovirus and rotavirus in Berlin, weeks 2011--W06 to 2017--W52. We revisit and extend two model variants specified in \cite{Held2017} for district-level norovirus data (there referred to as Models 7 and 8, Section 3.3). To model incidence in the twelve districts, both original models combine the formulation \eqref{eq:hhh4_mean_structure} and power law weights \eqref{eq:power_law} to describe spatial spread. The overdispersion parameter $\psi$ is shared across districts. The first model, which we refer to as the \textit{full model}, is defined as
\begin{align}
\log(\nu_{it}) & = \alpha^{(\nu)}_{i} + \beta^{(\nu)} x_t + \gamma^{(\nu)}\sin(2\pi t/\omega) + \delta^{(\nu)}\cos(2\pi t/\omega)\label{eq:nu_noro}\\
\log(\phi_{it}) & = \alpha^{(\phi)}_{i} + \gamma^{(\phi)}\sin(2\pi t/\omega) + \delta^{(\phi)}\cos(2\pi t/\omega),\label{eq:phi_noro}
\end{align}
where again $\omega = 52$. It features shared terms $\beta^{(\nu)}, \gamma^{(\cdot)}$ and $\delta^{(\cdot)}$ for seasonality, but district-specific intercepts $\alpha^{(\cdot)}_i$ in both the endemic and the epidemic parameters. As in \citet{Held2017}, the indicator $x_t$ for calendar weeks 52 and 1 aims to capture changes in reporting
behaviour or social contact patterns during the Christmas break, as quantified by the coefficients $\beta^{(\nu)}$. Note that the same indicator has also been included into the GLARMA reference model.

The second model is more parsimonious and features a \textit{gravity model} component \citep{Xia2004} instead of district-specific intercepts in the epidemic component. Equation \eqref{eq:phi_noro} is thus replaced by
\begin{align}
\log(\phi_{it}) & = \alpha^{(\phi)} + \tau^{(\phi)} \log(e_i) + \gamma^{(\phi)}\sin(2\pi t/\omega) + \delta^{(\phi)}\cos(2\pi t/\omega),
\end{align}
where $e_i$ is the fraction of the total population living in district $i$. The
amount of disease transmission thus scales with the population size of the ‘importing’ district \citep{Meyer2017}.

As mentioned in Section \ref{subsec:case_study_berlin}, epidemiological knowledge  on both diseases suggests that serial intervals may exceed one week. Also, it is known that gastrointestinal diseases are subject to considerable underreporting \citep{Tam2012} as well as meteorological influences \citep{Patel2013}. Consequently, we expect dependencies at higher lags and  replace formulation \eqref{eq:hhh4_mean_structure} by the more flexible \eqref{eq:multiv_distr_lag2}. Again, we apply nonparametric estimation of the lag weights and the three parametric versions \eqref{eq:pois_weights}--\eqref{eq:geom_weights}. For comparison we apply the first-order formulation \eqref{eq:fixed} as well as the reference models from Section \ref{subsec:reference_models}.

\subsubsection{Model fit and residual correlation}
\label{subsec:model_fits}

We start by contrasting some features of the model fits with and without weighting of past incidence. As the available time series are shorter here we use the full data set to obtain more reliable estimates. For the parametric serial weights the fitting took around four seconds per model on a standard laptop, while the unrestricted nonparametric versions took one minute. We chose $p = 5$ as for both diseases and model versions (full and gravity) the AIC values for $p > 5$ stayed constant for all parametric versions and increased for the nonparametric one, see Figures in the Supplementary Material. We also show plots of the profile likelihoods for the three parametric weighting schemes. The triangular version shows several local maxima for rotavirus, indicating that optimization has to be done with care under this parameterization (we adopt a grid-based optimization in the following to avoid getting stuck in local optima).

Table \ref{tab:aic} indicates that for both norovirus and rotavirus the model performance improves when weights for past incidence are included, with the geometric version performing best. For both diseases, this extension leads to a stronger improvement than modelling cross-district dependencies more flexibly (when moving from the gravity to the full model). This emphasizes the importance of including the information contained in observations from two or more weeks back in time.

\begin{table}[h!]
\caption{Differences in AIC relative to the first-order version of the gravity model (negative values indicate improvement).
}
\label{tab:aic}
\center
\begin{tabular}{r @{\hskip 1cm} r r @{\hskip 2cm} r r}
\hline
  & \multicolumn{2}{l}{Norovirus} & \multicolumn{2}{l}{Rotavirus}\\
 \hline  Weighting & gravity  & full  & gravity  & full \\
 scheme & model & model & model & model  \\
 \hline
first-order &    0.0 &  -63.5 &    0.0 &  -35.9 \\ 
  Poisson & -120.7 & -171.3 &  -78.7 & -106.9 \\ 
  triangular & -118.0 & -168.1 &  -74.8 & -102.5 \\ 
  geometric & -125.9 & \textbf{-174.2} &  -93.3 & \textbf{-118.2} \\ 
  unrestricted & -123.0 & -169.6 &  -92.9 & -117.3 \\ 
   \hline

\end{tabular}
\end{table}

\begin{figure}[h!]
\includegraphics[width=\textwidth]{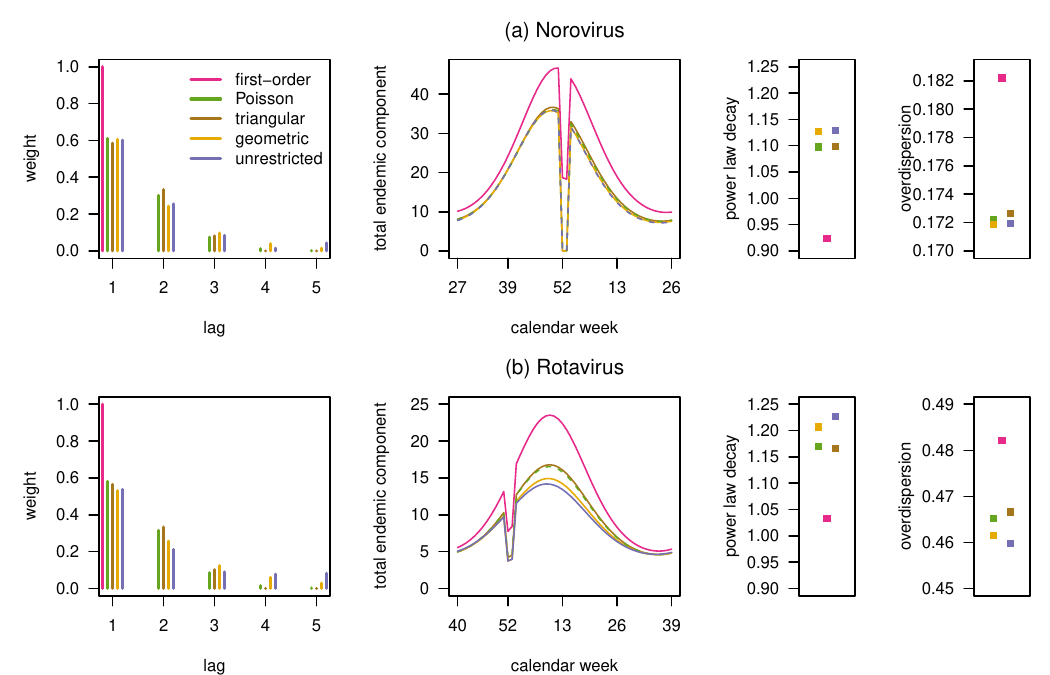} 

\caption{Estimated weights $\lfloor u_d\rfloor$, total endemic component $\sum_{i = 1}^{12} \nu_{it}$, decay parameter $\rho$ of the spatial power law and overdispersion parameter $\psi$ for the full model with different lag weighting schemes.}
\label{fig:lag_weights}
\end{figure}

In the remainder of this subsection we focus on the full model which showed better in-sample performance than the gravity model. Figure \ref{fig:lag_weights} shows selected features of models with different weight parameterizations. The weights $\lfloor u_d\rfloor$, shown in the first column, are similar for the two diseases, despite the differences in average serial intervals mentioned in Section \ref{subsec:case_study_berlin}. Under all four parameterizations, the first lag receives a weight slightly below 0.6. For both diseases the geometric weights mimic best the relatively slow decay in the unrestricted weights, explaining its good AIC values. As shown in the second column of Figure \ref{fig:lag_weights}, including the weighting schemes reduces the total endemic component $\sum_{i = 1}^{12}\nu_{it}$ considerably. A larger part of the incidence is thus explained by previous dynamics, which is a sign of improved model fit. Note that the lower values in calendar weeks $1$ and $52$ are due to the Christmas break. The power law decay parameters $\rho$ increase when lag weights are included (third column of Figure \ref{fig:lag_weights}), meaning that the model borrows slightly less information across regions. Finally, the overdispersion parameters $\psi$ are lower for the extended models, indicating less unexplained variability (fourth column).

The first-order models lead to some significant residual autocorrelations at lags two and three (plots shown in the Supplementary Material). These largely disappear in the extended models. 

\subsubsection{Predictive model assessment}
\label{subsec:predictive_asessement_berlin}

We assess the predictive performance of the different model variants for the period 2015-W01--2017-W52. Again, we re-fitted models each week including the most recent data. However, we chose the order $p$ only once based on the training data (2011-2014, 208 weeks). As for the models fitted to the full seven years of data in Section \ref{subsec:model_fits} this led to $p = 5$ for all parametric versions and the nonparametric version for rotavirus. For the norovirus model with nonparametric lag weights, however, $p = 3$ resulted in the best AIC on the training data, so that we used this value.

Table \ref{tab:logs} shows the average standardized multivariate log scores of one-week ahead forecasts for the norovirus and rotavirus time series (over the 156 weeks of test data). For norovirus, the additional flexibility of the full model seems to lead to somewhat improved predictive performance, but permutation tests indicate only weak evidence for an actual difference ($p$-values between 0.097 and 0.19 for the different parameterizations). For rotavirus both  model variants perform very similarly ($p$-values $\geq$ 0.57).

\begin{table}[h!]
\caption{Differences in mean standardized log scores of one-week-ahead forecasts over years 2015--2017, relative to the first-order gravity model (negative values indicate improvement). The two last rows provide results for the reference models.}
\label{tab:logs}
\center
\begin{tabular}{r @{\hskip 1cm} r r @{\hskip 2cm} r r}
\hline
  & \multicolumn{2}{l}{Norovirus} & \multicolumn{2}{l}{Rotavirus}\\
 \hline  Weighting & gravity  & full  & gravity  & full \\
 scheme & model & model & model & model  \\
 \hline
first-order &  0.000 & -0.006 &  0.000 &  0.003 \\ 
  Poisson & -0.015 & -0.021 & -0.017 & -0.015 \\ 
  triangular & -0.014 & -0.020 & -0.012 & -0.014 \\ 
  geometric & -0.017 & \textbf{-0.022} & \textbf{-0.019} & -0.018 \\ 
  unrestricted & -0.015 & -0.020 & -0.016 & -0.014 \\ 
   \hline
GLARMA &  &  0.000 &  &  0.011 \\ 
  naive &  &  0.091 &  &  0.090 \\ 
   \hline

\end{tabular}
\end{table} 

The inclusion of the lag weighting schemes, on the other hand, very consistently improves the predictive performance, with the geometric version yielding the best results. There is strong evidence that the latter predicts better than the respective first-order version for both model variants and diseases ($p$-values of around 0.001). However, there is only weak evidence for a performance difference between the geometric and the other other weighting schemes ($p$-values between 0.07 and 0.09 and between 0.07 and 0.11 for the full norovirus and rotavirus models, respectively).

All EE models clearly outperform the na\"ive seasonal models. The performance of the GLARMA model is similar to that of the first-order gravity model for norovirus and somewhat worse for rotavirus. Permutation tests indicate that the GLARMA approach is outperformed by the higher-order EE models ($p$-values of 0.002 or below when compared to the respective full geometric weight model).

\begin{figure}[htb]
\includegraphics[width=\textwidth]{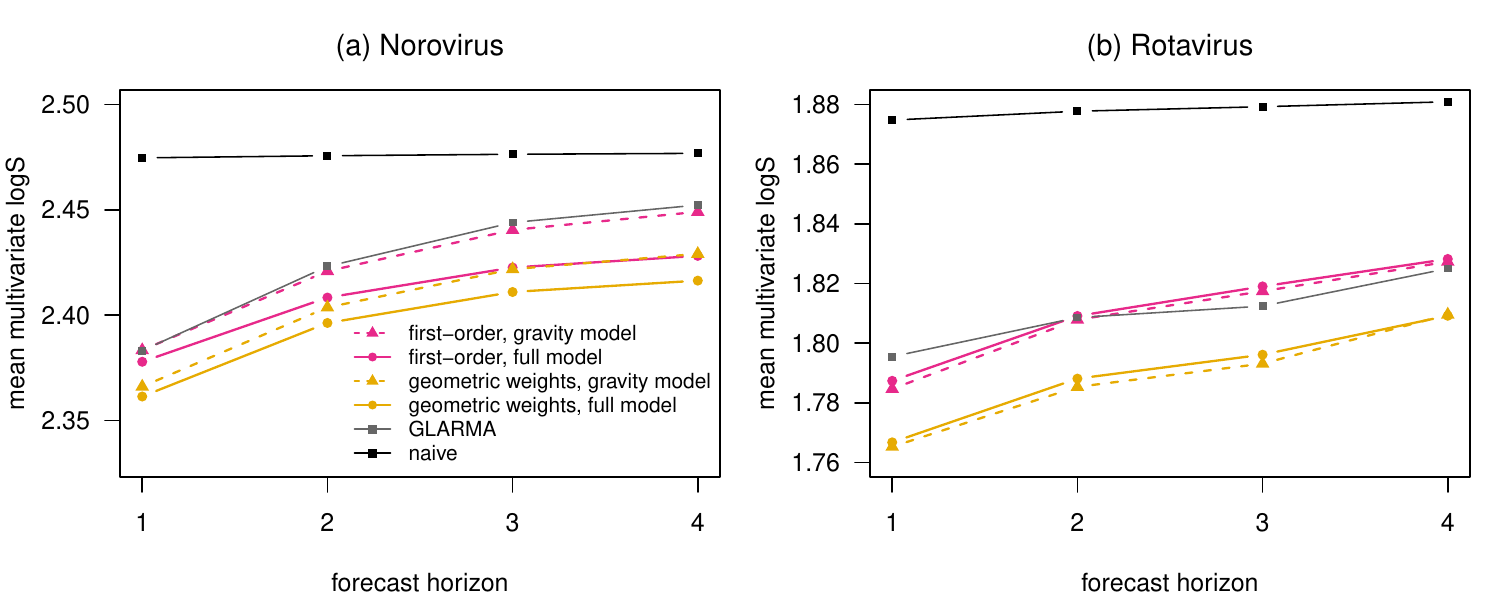} 

\caption{Average standardized multivariate log scores for norovirus and rotavirus forecasts at horizons $h = 1, 2, 3, 4$ weeks. Results are shown for first-order models and models with geometric weights. The na\"ive seasonal and the GLARMA model are shown for comparison.}
\label{fig:logS_horizons_noro_rota}
\end{figure}

Finally we compare the forecast performance of first-order EE models and models with geometric lag weights for horizons up to four weeks. Average standardized multivariate log scores over the test period are shown in Figure \ref{fig:logS_horizons_noro_rota}. For norovirus, the full model performs better than the gravity model also for horizons $h \geq 2$ ($p$-values of 0.062 and below), while for rotavirus the differences are again negligible. For both diseases and model versions, permutation tests indicate strong evidence that the geometric parameterization improves prediction at horizons $h = 2, 3, 4$ ($p$-values ranging from $< 0.0001$ to 0.007 for the two diseases and model versions). For norovirus the improvement corresponds roughly to the difference between two- and three-week-ahead forecasts. For rotavirus, it is even more pronounced as it corresponds to the difference between one- and three-week-ahead forecasts. Also at higher horizons, the GLARMA model yields similar performance as the first-order gravity model, while the na\"ive model performs considerably worse than all other approaches.

\section{Discussion}
\label{sec:discussion}

In this article we have extended the endemic-epidemic framework
to include weighting schemes for past incidence which are inspired by discrete-time serial interval distributions. In both case
studies the proposed parameterizations led to considerably improved
model fits, with the geometric scheme performing best. These
improvements also translated to improved predictive performance at
various forecasting horizons. There was no additional
benefit in estimating the lag weights in an
unrestricted nonparametric manner. The parametric formulations are
therefore an attractive option for practical analyses and
forecasts. As optimization of the log-likelihood function turned out
to be more difficult for the triangular weights and performance
tended to be slightly weaker, we consider the geometric and shifted
Poisson versions preferable.

Our model extension is motivated from serial interval distributions, but
some remarks are warranted. Firstly, as mentioned before, the weights $u_d$ are not only shaped by the serial interval distribution, but also other factors. These have to be taken into account if the purpose of an analysis is estimation of serial intervals rather than prediction. Moreover, to obtain valid estimates of transmission parameters, the observation interval needs to be chosen in a suitable way \citep{Nishiura2010, Reich2016}, ideally similar or shorter than the serial interval. For prediction purposes, however, this is not crucial, as the EE model can adapt to the increased dispersion resulting from within-week dynamics via its overdispersion parameter. We therefore think that our model extension is applicable to short-term forecasting of a wide range of diseases with short to medium-length serial intervals.

A property of the proposed formulation
is that the weighting parameter $\kappa$  is assumed to be time-constant and identical across units, which is motivated from the assumption of time-invariant serial interval distributions. This may be restrictive in some applications, for instance due to time-varying social contact patterns, meteorological conditions and dominant strains. However, while in principle it would be possible to allow for a time-varying $\kappa_t$, we expect identifiability to be poor. Sometimes the EE framework is also applied to analyse two diseases jointly \citep{Paul2008, Chiavenna2019}, in which case using separate parameters $\kappa_i$ could be of interest. In spatio-temporal settings as in our work, however, we prefer to use one shared parameter $\kappa$.

In the present article we only assessed the forecasting performance for $h$-week ahead
forecasts. Other potential forecast targets include the entire incidence curve over
a season \citep{Held2017} or more specific features like the onset timing, peak timing
and peak incidence \citep{Ray2017, McGowan2019}. These, too, are expected to be dependent
across geographical sub-units. It would therefore be of interest to generate and evaluate forecasts
in a multivariate fashion as illustrated for $h$-week ahead forecasts in the present article.
But even if forecasts are only evaluated at a univariate aggregate level, initial modelling at a finer resolution
is a promising strategy which can improve forecasts \citep{Held2017}.

\appendix

\section{Software and reproducibility}
\label{app:software}

Software for model fitting and forecasting is provided in the \texttt{R} package \texttt{hhh4addon} which extends the functionality of the \texttt{surveillance} package \citep{Meyer2017}. It is available at \url{https://github.com/jbracher/hhh4addon}. The package also contains the data presented in Section \ref{sec:case_studies}, which were obtained from the web platform of the Robert Koch Institute ({\url{https://survstat.rki.de}}) and the Supplementary Material of \cite{Ray2017}. Codes to reproduce the presented analyses are available at \url{https://github.com/jbracher/dengue_noro_rota}.

\newpage


\end{document}